\documentclass[runningheads]{llncs}
\usepackage[T1]{fontenc}
\usepackage{graphicx}
\usepackage{amsthm}
\usepackage{amsmath,amsfonts}
\usepackage[utf8]{inputenc}
\usepackage{makecell} 
\usepackage{array}
\usepackage{soul}	 
\AtBeginDocument{%
  }

\newcommand{\sysname}{\text{PrivGNN}}
\usepackage{textcomp}
\usepackage{stfloats}
\usepackage{url}
\usepackage{verbatim}
\usepackage{graphicx}
\usepackage{cite}
\usepackage{marvosym}
\usepackage{ifsym}
\graphicspath{{Figs/}}
\usepackage[table]{xcolor}

\usepackage{mathtools}
\usepackage{IEEEtrantools} 
\usepackage{hyperref}
\hypersetup{colorlinks=true,linkcolor=blue,anchorcolor=blue,citecolor=blue}

\usepackage{graphicx,color}
\usepackage{subfigure}
\graphicspath{{fig/}}
\usepackage{pgfplots}
\pgfplotsset{compat=1.14}
\newlength\figwidth
\usepackage{threeparttable}
\usepackage{multirow, booktabs, diagbox}
\usepackage{tabularx} 
\usepackage{mathtools}
\usepackage{adjustbox}

\usepackage{algorithm}
\usepackage{algorithmicx}
\usepackage{algpseudocode}
\newcommand*{\circled}[1]{\lower.7ex\hbox{\tikz\draw (0pt, 0pt)%
    circle (.4em) node {\makebox[0.4em][c]{\small #1}};}}

\usepackage{indentfirst}
\usepackage[utf8]{inputenc}
\usepackage{makecell}
\usepackage{textcomp,booktabs}
\usepackage{enumitem}
\newlist{todolist}{itemize}{2}
\setlist[todolist]{label=$\square$}
\usepackage{pifont}
\usepackage{inconsolata}
\usepackage{anyfontsize}
\newcommand{\cxmark}{\textcolor{black}{\ding{51}}\textsuperscript{\textcolor{black}{\kern-0.45em\scriptsize\ding{55}}}}

\usepackage{tikz}
 
\usetikzlibrary{fit,calc}

\colorlet{pink}{red!20} 
\definecolor{mycolor}{RGB}{232,247,253}

\usepackage{url}
\usepackage{setspace}

\usepackage[most]{tcolorbox}

\tcbuselibrary{most}
\newtcolorbox{mybox}[2][]{%
  enhanced,
  title        = {#2},
  attach boxed title to top left={xshift=+3mm,yshift*=-3mm},
  colback      = white,
  colframe     = black,
  fonttitle    = \bfseries,
  fontupper    = \small,
  fontlower    = \small,
  colbacktitle = black!3!white,
  coltitle     = black,
  #1
}

\newtcbox{\xmybox}[1][red]{on line, arc=7pt,colback=#1!10!white,colframe=#1!50!black, before upper={\rule[-3pt]{0pt}{10pt}},boxrule=1pt, boxsep=0pt,left=6pt,right=6pt,top=1pt,bottom=1pt}

\begin{document}
\title{$\sysname$: High-Performance Secure Inference for Cryptographic Graph Neural Networks}

\author{Fuyi Wang\inst{1,2}  \and
Zekai Chen\inst{3}\and
Mingyuan Fan\inst{4} \and
Jianying Zhou\inst{2}\and
Lei Pan\inst{1} \and
Leo Yu Zhang\inst{5}\textsuperscript{\Letter} }
\authorrunning{F. Wang et al.}

\institute{Deakin University, Geelong, Australia \and
Singapore University of Technology and Design, Singapore\\ \and
Fuzhou University, Fuzhou, China\\ \and
East China Normal University, Shanghai, China\\ \and
Griffith University, Gold Coast, Australia\\ \email{leo.zhang@griffith.edu.au} }

%
%
%
%
\maketitle              
\begin{abstract} 
Graph neural networks (GNNs) are powerful tools for analyzing and learning from graph-structured (GS) data, facilitating a wide range of services. Deploying such services in privacy-critical cloud environments necessitates the development of secure inference (SI) protocols that safeguard sensitive GS data. However, existing SI solutions largely focus on convolutional models for image and text data, leaving the challenge of securing GNNs and GS data relatively underexplored. In this work, we design, implement, and evaluate $\sysname$, a lightweight cryptographic scheme for graph-centric inference in the cloud. By hybridizing additive and function secret sharings within secure two-party computation (2PC), $\sysname$ is carefully designed based on a series of novel 2PC interactive protocols that achieve $1.5\times \sim 1.7\times$ speedups for linear layers and $2\times \sim 15\times$ for non-linear layers over state-of-the-art (SotA) solutions. A thorough theoretical analysis is provided to prove $\sysname$'s correctness, security, and lightweight nature. Extensive experiments across four datasets demonstrate $\sysname$'s superior efficiency with $1.3\times \sim 4.7\times$ faster secure predictions while maintaining accuracy comparable to plaintext graph property inference. 

\end{abstract}

\section{Introduction}
\label{sec:Introduction}
\vspace{-0.6\baselineskip}
Recently, the rapid development of graph neural network (GNN) techniques has significantly impacted various domains such as drug discovery \cite{liao2022secmpnn}, social networks \cite{wang2023secgnn}, and recommendation systems \cite{liao2023ppgencdr}.  
For example, pharmaceutical enterprises can train the specific GNNs \cite{gilmer2017neural} on known drug compounds to predict the binding potential of new molecules with target proteins, such as those related to cancer. 
These well-trained GNNs can then be offered as AI-driven services on a pay-as-you-go basis, allowing end-users to enhance drug discovery or analyze complex molecular data.
However, graph-structured (GS) data often represents sensitive and proprietary information, making end-users hesitant to outsource it to service providers due to privacy concerns. 
Also, highly specialized GNNs for graph-centric services are considered valuable intellectual property and should be protected against leakage while circumventing reverse engineering risks \cite{wu2022linkteller}.

In response to these concerns, privacy-preserving deep learning (PPDL) sche- mes have been developed to enable secure inference (SI) using cryptographic techniques such as homomorphic encryption (HE) \cite{paillier1999public} and/or multi-party computation (MPC) \cite{goldreich2009foundations}. However, HE-based schemes often suffer from heavy computational and communication overhead due to inefficient exponentiation and ciphertext expansion. As a result, recent efforts focus on leveraging MPC \cite{goldreich2009foundations} for SI with secret-shared neural networks (NNs) and inputs, offering a more efficient alternative \cite{makri2021rabbit}.
Advanced SI solutions \cite{rathee2020cryptflow2,huang2022cheetah,yang2023fssnn}, have successfully secured convolutional neural networks (CNNs) for unstructured data (e.g., images and text). 
However, SI of GNNs in the cloud for complex GS data, while explored in some initial studies \cite{ran2022cryptogcn,wang2023secgnn,xu2024oblivgnn,peng2024lingcn}, remains relatively underdeveloped. 

\textbf{Related works and challenges.} Existing cryptographic GNN approaches primarily introduce HE \cite{ran2022cryptogcn,peng2024lingcn} and MPC \cite{wang2023secgnn} to secure node features in graph data. However, they do not fully protect the relational structure of graphs, such as the maximum node degree \cite{wang2023secgnn} and the range of node degrees \cite{ran2022cryptogcn} (\textbf{Challenge \circled{1}}). For instance, while SecGNN \cite{wang2023secgnn} optimizes node representation by reducing the adjacency matrix size from $\mathcal{O}(N^2)$ to $\mathcal{O}(N \cdot d_{max})$---resulting in over 90\% reduction for real-world datasets---it leaks the maximum degree $d_{max}$ in the graph.
Moreover, these approaches \cite{ran2022cryptogcn,peng2024lingcn,wang2023secgnn} often rely on cryptographic primitives that are computationally and communication-intensive, particularly for non-linear operations (\textbf{Challenge \circled{2}}). 
Non-linear operations, such as ReLU, implemented with primitives like garbled circuits in MPC or HE, are several orders of magnitude more expensive than linear layers in computation and communication \cite{zhou2023bicoptor}. 
While some optimizations have been proposed for non-linear operations \cite{rathee2020cryptflow2, huang2022cheetah, zhou2023bicoptor}, these approaches often impose heavy online computational burdens or require multiple communication rounds. Such constraints pose challenges to scalability and efficiency for real-world graph data containing millions or even billions of nodes and edges.
Lastly, recent efforts like SecMPNN \cite{liao2022secmpnn} and OblivGNN \cite{xu2024oblivgnn} optimize non-linear operations via lightweight secret sharing primitives in a fully outsourced setting, where a group of cloud servers jointly perform SI over encrypted/shared GNNs and inputs. However, this setup raises concerns about cloud-server collusion, relying on a strong non-collusion assumption for security models (\textbf{Challenge \circled{3}}). SecMPNN \cite{liao2022secmpnn} also requires continuous third-party assistance during online phases, adding additional complexity.

To tackle the above challenges, we design, implement, and evaluate $\sysname$, a high-performance cryptographic scheme for secure GNN inference with a client-server setup.
In this secure two-party computation (2PC) setup, $\sysname$ distributes the SI workload between the client and the service provider, mitigating cloud-collusion risks and enhancing security. 
To improve efficiency, $\sysname$ employs lightweight additive secret sharing (AddSS) and function secret sharing (FuncSS) in an offline-online paradigm, significantly reducing the computational overhead during the online phase. By carefully designing the offline phase, the online phase is reduced to only one round of interaction with lightweight computation. 
Specifically, we develop and optimize a line of secure protocols for both linear and non-linear computations. 
These protocols integrate seamlessly into various layers across various graph-centric services, reducing the online SI latency.
In summary, our contributions are threefold. 

\begin{itemize}[leftmargin=*,topsep=0pt] 
    \item[-] We present $\sysname$, a fast SI scheme for graph-centric services via the delicate synergy of GNNs and cryptography.
    $\sysname$ employs lightweight AddSS to securely protect the graph structure, node features, and well-trained GNNs, without relying on a strong non-collusion assumption in the 2PC setup. 
    \item[-] With AddSS and FuncSS, we propose a set of secure protocols for multiplication, comparison-based activations, and sigmoidal activation variants. 
    Our designs shift the computational-intensive operations to the offline phase and streamline online communication with only one-round interaction. 
    \item[-] We formally prove the correctness, efficiency, and security of $\sysname$. We conduct extensive experiments showing that our protocols outperform SotA works. Results on various datasets highlight $\sysname$'s efficiency and scalability in applications like image classification and molecular property recognition.
\end{itemize}

\section{Preliminaries}
\label{sec:Preliminaries}
\vspace{-0.6\baselineskip}
\textbf{Additive secret sharing} (AddSS) \cite{demmler2015aby} divides a secret message over the plaintext space into multiple shares, each of which is distributed to different parties. AddSS's key property is that the shares can be combined (added) together to reconstruct the original secret, while no individual shareholder possesses adequate information to determine the secret message independently. This paper adopts the 2-out-of-2 AddSS over the ring $\mathbb{Z}_{2^{l}}$ and the definition is given below.
\vspace{-2mm}
\begin{definition}
    A 2-out-of-2 AddSS scheme over the ring $\mathbb{Z}_{2^{l}}$ is a pair of probabilistic polynomial-time (PPT) algorithms $\{\texttt{Share}, \texttt{Reconstruct}\}$ where
\begin{itemize}[leftmargin=*,topsep=0pt]
    \item[-]  $\texttt{Share}$: On the secret message $x \in \mathbb{Z}_{2^{l}}$, $\texttt{Share}$ outputs shares $\{\langle x \rangle_0, \langle x \rangle_1\} \in \mathbb{Z}_{2^{l}}$, s.j. $x=\langle x \rangle_0 + \langle x \rangle_1 \pmod{2^l}$.
    \item[-]  $\texttt{Reconstruct}$: With $2$ shares $\{\langle x \rangle_0, \langle x \rangle_1\} \in \mathbb{Z}_{2^{l}}$, $\texttt{Reconstruct}$ outputs $x$ over the plaintext space $\mathbb{Z}$.
\end{itemize}
\end{definition}

\vspace{-2mm}
In the case of $l >1$ (e.g., $l =32$) which supports arithmetic operations (e.g., addition and multiplication), the arithmetic share pair is denoted by $\langle \cdot \rangle_ \gamma $ ($\gamma \in \{0,1\}$). In the case of $l =1$ which supports Boolean operations XOR $(\oplus)$, NOT ($\neg$) and AND ($\otimes$), the Boolean share pair is denoted by $\left [ \cdot \right ]_ \gamma$. In the following, we assume all arithmetic operations to be performed in the ring $\mathbb{Z}_{2^{l}}$ (i.e., all arithmetic operations are mod $2^l$).

\textbf{Function secret sharing} (FuncSS) \cite{boyle2015function,boyle2021function} within the 2PC setup divides a function $f:\mathbb{G}_{\textnormal{in}} \rightarrow \mathbb{G}_{\textnormal{out}}$ into 2 shares $\{f_0,f_1\}$, where $\mathbb{G}_{\textnormal{in}}$ and $\mathbb{G}_{\textnormal{out}}$ are input and output groups. Each party receives one of the function shares, and for any input $x$, there exists $f_0(x)+f_1(x) = f(x)$. The definition of FuncSS is given below.

\begin{definition}
    A 2PC FuncSS scheme over the ring $\mathbb{Z}_{2^{l}}$ is a pair of algorithms $\{\texttt{Gen}, \texttt{Eval}\}$ where
\begin{itemize}[leftmargin=*,topsep=0pt]
    \item[-] $\texttt{Gen}$: With the security parameter $\kappa$ and a function $f$, the PPT key generation algorithm $\texttt{Gen}(1^{\kappa},f)$ outputs a pair of keys $\{k_0, k_1\}$, where each key implicitly represents $f_{\gamma}:\mathbb{G}_{\textnormal{in}} \rightarrow \mathbb{G}_{\textnormal{out}}$. 
    
    \item[-] $\texttt{Eval}$: With the party identifier $\gamma \in \{0,1\}$, the key $k_{\gamma}$ (key defining $f_{\gamma}$), and the public input $x \in \mathbb{Z}_{2^{l}}$, the PPT evaluation algorithm 
    $\texttt{Eval}(\gamma,k_{\gamma},x)$ outputs $y_{\gamma} \in \mathbb{Z}_{2^{l}}$, i.e., the value of $f_{\gamma}(x)$, where $f(x)= \sum_{\gamma=0}^{1}y_{\gamma}$.
\end{itemize}
\end{definition}

\vspace{-0.4\baselineskip} 
\section{System Overview}
\label{sec:SystemOverview}
\begin{figure}[!t]
\centering
\centerline{\includegraphics[width=0.9\textwidth]{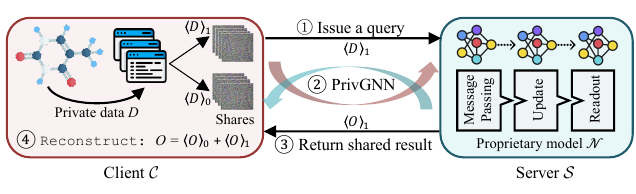}}
\vspace{-4mm}
\caption{{A client-server secure inference scenario.}} 
\vspace{-4mm}
\label{fig:system} 
\end{figure}

\vspace{-0.6\baselineskip}
\textbf{System model.} We consider the SI scenario with a client-server setup, where the client $\mathcal{C}$ (e.g., a drug laboratory) holds private GS data $D$, and the server $\mathcal{S}$ possesses a well-trained GNN model $\mathcal{N}$ with private weights $W$, illustrated in Fig.~\ref{fig:system}. 
$\mathcal{C}$ intends to utilize the model $\mathcal{N}(W, \cdot)$ to facilitate accurate results on its data $D$ (i.e., $\mathcal{N}(W, D)$), while ensuring that $\mathcal{C}$'s data remains confidential. We regard $\gamma \in \{0,1\}$ as the identifier of a party, $\gamma =0 $ represents $\mathcal{C}$ and $\mathcal{S}$ otherwise. 
Specifically, \circled{1} To ensure the privacy of raw graph ${D}$, $\mathcal{C}$ uses AddSS to split $D$ into two shares (i.e., $D = \left \langle D \right \rangle_0 + \left \langle D \right \rangle_1$), then sends the share $\left \langle D \right \rangle_1$ to $\mathcal{S}$ to issue a SI query. \circled{2} $\mathcal{C}$ and $\mathcal{S}$ collaboratively take charge of SI tasks via utilizing a series of secure computation protocols in $\sysname$. \circled{3} After execution of $\sysname$, $\mathcal{S}$ returns the shared inference result $\left< {O} \right>_1$ to $\mathcal{C}$. \circled{4} Upon receiving $\left< {O} \right>_1$, the client $\mathcal{C}$ reconstructs to get the final plaintext result via $O=\left \langle O \right \rangle_0 + \left \langle O \right \rangle_1$ to complete the prediction and inference services.

\textbf{Threat model.} We assume that $\mathcal{C}$ and $\mathcal{S}$ in $\sysname$ are semi-honest, i.e., they honestly obey the specification of the protocols, yet attempt to learn auxiliary information from intermediate shares. Non-collusion is unnecessary to assume, as collusion would imply a willingness to disclose their private data. The service provider (i.e., $\mathcal{S}$), operated by reputable vendors like Google, is incentivized to follow the protocols, as poor performance or violations of privacy regulations would have an irreversible negative impact on their credibility and profitability.
Therefore, the semi-honest assumption is practical, as witnessed in existing secure outsourcing works \cite{ohata2020communication,rathee2020cryptflow2,cheng2023private,liao2022secmpnn,wang2023secgnn}. 
We define security in the real/ideal world. In the real world, protocol $\prod$ is executed, yielding the output $\textsf{Real}^{\prod}_{\mathcal{A}}(1^{\kappa})$, while in the ideal world, the ideal functionality $\mathcal{F}$ is executed by a trusted third party, yielding the output $\textsf{Ideal}^{\mathcal{F}}_{\textsf{Sim}}(1^{\kappa})$. Here, $\mathcal{A}$ is a stateful adversary, $\textsf{Sim}$ is a stateful simulator, and $\kappa \in \mathbb{N}^{+}$ is the security parameter.
 
\begin{definition} 
    Let $\mathcal{F}=\{\mathcal{F}_0, \mathcal{F}_1\}$ be a ideal-world function and $\prod$ be a real-world protocol computing $\mathcal{F}$.  For any PPT adversary $\mathcal{A}$ in the real world, there exists a PPT simulator $\textsf{Sim}$ in the ideal world, such that for corrupted party $\mathcal{P} \subset  \{\mathcal{S}, \mathcal{C}\}$
    \vspace{-2mm}
    \begin{IEEEeqnarray}{rCL}
    \textsf{Ideal}^{\mathcal{F}}_{\textsf{Sim}_{\mathcal{P}}}(1^{\kappa}, D, W)\equiv_{c} \textsf{Real}^{\prod}_{\mathcal{A}_{\mathcal{P}}}(1^{\kappa}, D, W).
    \label{security}
    \vspace{-2mm}
    \end{IEEEeqnarray} 
\noindent where $\equiv _{c}$ denotes computational indistinguishability against PPT adversaries except for a negligible advantage.
\end{definition}

\vspace{-0.4\baselineskip}
\section{Our Approach}
\label{sec:Supporting Protocols}
\vspace{-0.6\baselineskip}
$\sysname$ introduces the offline-online paradigm to minimize the online communication rounds and computational costs. 
This paradigm alleviates the computational burden on resource-constrained clients during the online phase, effectively enhancing the client experience.
To achieve lightweight online performance, we use AddSS and FuncSS to design multiplication, quadratic polynomial, comparison protocols. 
In these protocols, the computational-intensive cryptographic operations, i.e., $\texttt{Gen}^{\text{BV}}$ for generating Beaver multiplication triples \cite{beaver1992efficient} and $\texttt{Gen}^{\text{key}}$ for generating FuncSS keys \cite{boyle2015function}, are performed during the data-independent offline phase. These operations can be implemented using either a trusted third party \cite{hao2023fastsecnet,boyle2021function, yang2023fssnn} or specific 2PC protocols \cite{mishra2020delphi,cheng2023private}.
This paper opts for the former one. All these protocols integrate seamlessly into NN layers. 

\vspace{-0.4\baselineskip}
\subsection{Secure Matrix Multiplication Protocol} \label{ssub:Secure Matrix Multiplication Protocol}
\vspace{-0.2\baselineskip}
Given the secret-shared $\langle X \rangle$ and the server's plaintext $Y$ for secure matrix multiplication protocol $\mathtt{SMatMul}$ (i.e., $\langle Z\rangle=\langle X\times Y \rangle$), the key insight is that the output $\langle Z\rangle$ and client's input $\langle X \rangle$ are secret-shared, while variable $Y$ is the plaintext held by the server $\mathcal{S}$. We reformulate $X\times Y \rightarrow (X-A)\times Y+ A\times Y$, where $A$ is randomness held by the client $\mathcal{C}$. Thus, $\mathtt{SMatMul}$'s offline phase focuses on pre-computing the shares of $A\times Y$, and then the online phase evaluates $(X-A)\times Y$ in plaintext by $\mathcal{S}$. 
Compared with \cite{mishra2020delphi,hao2023fastsecnet}, $\mathtt{SMatMul}$  reduces \textit{one} unnecessary auxiliary randomness $r$, eliminating $r$-related offline computation and minimizing pseudorandom generator usage. 
Algorithm~\ref{alg:MatMul} presents the main steps and the details of $\mathtt{SMatMul}$ are described below.

\begin{itemize}[leftmargin=*,topsep=0pt]
    \item[-] The offline phase aims to learn the shared $A\times Y$. Specifically, client and server sample $A$ and $B$ randomly, respectively, and then jointly invoke $\texttt{Gen}^{\text{BV}}(A,B)$ to generate matrix multiplication triples $\{A, \langle C\rangle_0 \}$ for client and $\{B, \langle C \rangle_1 \}$ for server, where $\langle C\rangle_0+ \langle C\rangle_1=A \times B$. After that, the server sends $Y-B$ to the client. Next, the client locally compute $\langle A\times Y\rangle_0=A\times(Y-B)+\langle C\rangle_0$ and the server sets $\langle A\times Y\rangle_1=\langle C\rangle_1$.
    \item[-] The online phase aims to learn the final shared $X\times Y$. The client computes $\langle X \rangle_0-A$ and sends its result to the server. Finally, the client holds $\langle X\times Y\rangle_0=\langle A\times Y\rangle_0$ directly and the server learns $\langle X\times Y\rangle_1=(\langle X \rangle_0+\langle X \rangle_1-A)\times Y+\langle C\rangle_1$.
\end{itemize}

\begin{algorithm}[!t]
\caption{\label{alg:MatMul} {Secure Matrix Multiplication Protocol} }
\textbf{Input:} $\mathcal{C}$ holds $\langle X \rangle_0 \in \mathbb{Z}_{2^{l}}^{m_1 \times m_2}$. $\mathcal{S}$ holds $\langle X \rangle_1 \in \mathbb{Z}_{2^{l}}^{m_1 \times m_2}$, $Y \in \mathbb{Z}_{2^{l}}^{m_2 \times m_3}$.
\par
\textbf{Output:} $\mathcal{C}$ learns $\langle Z \rangle_0 \in \mathbb{Z}_{2^{l}}^{m_1 \times m_3}$ . $\mathcal{S}$ learns $\langle Z \rangle_1 \in \mathbb{Z}_{2^{l}}^{m_1 \times m_3}$, where $Z = X \times  Y$.
\begin{algorithmic}[1]
\Statex \textbf{\# Offline Phase:} $\langle A\times Y \rangle$
\State $\mathcal{C}$ and $\mathcal{S}$ sample $A \stackrel{\$}{\leftarrow} \mathbb{Z}_{2^{l}}^{m_1 \times m_2},B \stackrel{\$}{\leftarrow} \mathbb{Z}_{2^{l}}^{m_2 \times m_3}$, and then jointly invoke $\texttt{Gen}^{\text{BV}}(A,B) \rightarrow \{\langle C\rangle_0, \langle C\rangle_1 \}$, where $C = A \times B$.
\State $\mathcal{S}$ computes $Y-B$ and sends its result to $\mathcal{C}$ and sets $\langle A\times Y\rangle_1=\langle C\rangle_1$.
\State $\mathcal{C}$ computes $\langle A\times Y\rangle_0=A\times(Y-B)+\langle C\rangle_0$. 
\Statex \textbf{\# Online Phase:} $\langle X\times Y\rangle$
\State $\mathcal{C}$ computes $\langle X \rangle_0-A$ and sends its result to $\mathcal{S}$. $\mathcal{C}$ sets $\langle Z \rangle_0=\langle A\times Y\rangle_0$.
\State $\mathcal{S}$ computes $\langle Z \rangle_1= (X-A)\times Y+\langle C\rangle_1=(\langle X \rangle_0+\langle X \rangle_1-A)\times Y+\langle C\rangle_1$.
\end{algorithmic}
\end{algorithm}

\textbf{Secure element-wise multiplication protocol.}
Algorithm~\ref{alg:MatMul} is applicable for secure element-wise multiplication protocol $\mathtt{SEleMul}$, with the modification of replacing matrix multiplication $\times$ with element-wise multiplication $\odot$, while ensuring that the sizes of $X$, $Y$, $A$, and $B$ remain the same:
$\left<Z\right>=\mathtt{SEleMul}(\left<X\right>,Y)$, where $\left<z_{i,j}\right>=\mathtt{SMatMul}(\left<x_{i,j}\right>,y_{i,j})$.

 \textbf{Secure fully-connected and convolutional layers.} A fully-connected layer $\textsf{SecFC}$ in NNs is essentially a matrix multiplication, thus $\textsf{SecFC}$ is implemented directly using $\mathtt{SMatMul}$: $\textsf{SecFC}(\left<D\right>,W) =\mathtt{SMatMul}(\left<D\right>,W)$. 
A convolutional layer $\textsf{SecCONV}$ can be expressed as a matrix multiplication with the help of reshape techniques~\cite{kumar2020cryptflow}: $\textsf{SecCONV}(\left<D\right>,W)$ = $\texttt{ReshapeOutput}(\mathtt{SMatMul}(\texttt{Reshape}$ $(\texttt{Input}{(\left<D\right>)},\texttt{ReshapeFilter}{(W)}))$.


\vspace{-0.4\baselineskip}
\subsection{Secure Quadratic Polynomial Protocol} 
\label{ssub:Secure Quadratic Polynomial Protocol}
\vspace{-0.4\baselineskip}
We initially propose the secure quadratic polynomial protocol $\mathtt{SQuaPol}$, which computes $z = p_2 x^2 + p_1 x + p_0$. In this protocol, both parties, $\mathcal{C}$ and $\mathcal{S}$, hold secret-shared inputs $\left<x\right>$ and outputs $\left<z\right>$, while $\mathcal{S}$ possesses the plaintext coefficients ${p_0, p_1, p_2}$. We then extend $\mathtt{SQuaPol}$ to support polynomials of arbitrary degrees.
Inspired by the customized $\mathtt{SMatMul}$ in Sec.~\ref{ssub:Secure Matrix Multiplication Protocol}, we have the insight of the following reformulation:

\vspace{-5mm}
\begin{IEEEeqnarray}{rCL}
z &=& p_2 x^2+ p_1 x+ p_0 = p_2 (x-a+a)^2+ p_1 (x-a+a)+ p_0 \qquad \triangleright f\leftarrow x-a   \nonumber \\
&=& p_2 (f+a)^2+ p_1 (f+a)+ p_0 =  p_2 f^2 + \underline{p_2 a^2} + {\underline{2p_2 a}f }+ p_1 f + \underline{p_1 a} + p_0, 
\label{eq:QuaPol}
\vspace{-1mm}
\end{IEEEeqnarray}


where $a$ is the randomness held by the client $\mathcal{C}$ which is independent of the input $x$. 
Hence, $\mathtt{SQuaPol}$ can be efficiently divided into online and offline phases.
The offline phase, which performs the computations labeled with underline in Eq.~\ref{eq:QuaPol}, is precomputed without knowing the input $x$, reducing the computational burden during the online phase.
In the online phase, the client and server only need to compute the shared value $f$ with minimal interaction, requiring just one round of communication and one message per party. This design significantly reduces both online communication overhead and latency, highlighting the efficiency and practicality of $\sysname$. 
Algorithm~\ref{alg:QuaPol} presents the details of $\mathtt{SQuaPol}$ and the main steps are given below.

\begin{itemize}[leftmargin=*,topsep=0pt]
    \item[-] The offline phase aims to learn the shared ${\langle p_2 a^2 \rangle_\gamma}, {\langle 2p_2 a \rangle_\gamma}, {\langle p_1 a\rangle_\gamma}$ ($\gamma \in \{0,1\}$) and precompute the parameter on the server side,  which is used in the online phase of $\langle 2p_2 a  f\rangle$ in Eq.~\ref{eq:QuaPol}. Therefore, four sets of multiplication triples need to be generated by invoking $\texttt{Gen}^{\text{BV}}$ four times. 
    \item[-] During the online phase, after one communication round, the client computes $\langle z\rangle_0 \leftarrow \langle p_2 a^2 \rangle _0 + 2(\underbrace{(a_4 (e_4 +f_5)+ \langle c_4\rangle_0)}_{\langle p_2af \rangle_0}) +\langle p_1 a \rangle _0 $ and the server computes $\langle z\rangle_1 \leftarrow p_2({\underbrace{f_5 +\langle x \rangle_1}_{f}})^2 +\langle p_2 a^2 \rangle_1+ 2(\underbrace{(f_5 +\langle x \rangle_1 )(f_4+\langle p_2 a \rangle_1)+\langle c_4\rangle_1}_{\langle p_2af \rangle_1}) + p_1 (\underbrace{f_5 +\langle x \rangle_1}_{f})+ \langle p_1 a \rangle _1+ p_0$.
\end{itemize}



\begin{algorithm}[!t]
\caption{\label{alg:QuaPol}{Secure Quadratic Polynomial Protocol} }
\textbf{Input:} $\mathcal{C}$ holds $\langle x \rangle_0 \in \mathbb{Z}_{2^{l}}$. $\mathcal{S}$ holds $\langle x \rangle_1 \in \mathbb{Z}_{2^{l}}$, $p_0, p_1, p_2 \in \mathbb{Z}_{2^{l}}$.
\par
\textbf{Output:} $\mathcal{C}$ learns $\langle z \rangle_0 $ . $\mathcal{S}$ learns $\langle z \rangle_1 $, where $z = p_2 x^2+ p_1 x+ p_0$.
\begin{algorithmic}[1]
\Statex \textbf{\# Offline Phase:} $\langle p_2a^2 \rangle$, $\langle p_2a \rangle$, $\langle p_1 a \rangle$,
\State $\mathcal{C}$ and $\mathcal{S}$ sample $\{a,a_1,a_2,a_3,a_4\} \stackrel{\$}{\leftarrow} \mathbb{Z}_{2^{l}}^5,\{b_1,b_2,b_3,b_4\}  \stackrel{\$}{\leftarrow} \mathbb{Z}_{2^{l}}^4$, and then jointly invoke $\texttt{Gen}^{\text{BV}}(a_i,b_i)$ ($\forall i \in \{1,2,3,4\}$) to generate triples $\{a_i, \langle c_i\rangle_0 \}$ for $\mathcal{C}$ and $\{b_i, \langle c_i\rangle_1 \}$ for $\mathcal{S}$, where $c_i = a_i b_i$.
\State $\mathcal{S}$ computes $e_1\leftarrow p_1 -b_1, e_2\leftarrow p_2 -b_2, e_3\leftarrow p_2 -b_3$, and sends $e=\{e_1, e_2, e_3\}$ to $\mathcal{C}$.   
\State $\mathcal{C}$ computes $\langle p_1 a \rangle _0 \leftarrow {a_1}{e_1}+\langle c_1\rangle_0, \langle p_2 a \rangle _0 \leftarrow {a_2}{e_2}+\langle c_2\rangle_0, \langle p_2 a^2 \rangle _0 \leftarrow {a_3}{e_3}+\langle c_3\rangle_0$, $f_1\leftarrow a -a_1, f_2\leftarrow a -a_2, f_3\leftarrow a^2 -a_3$, $f_4= \langle p_2 a \rangle _0 -{a_4}$, and sends $f=\{f_1, f_2, f_3, f_4\}$ to $\mathcal{S}$.
\State $\mathcal{S}$ computes $\langle p_1 a \rangle _1 \leftarrow {p_1}{f_1}+\langle c_1\rangle_1, \langle p_2 a \rangle _1 \leftarrow {p_2}{f_2}+\langle c_2\rangle_1, \langle p_2 a^2 \rangle _1 \leftarrow {p_2}{f_3}+\langle c_3\rangle_1$.
\Statex \textbf{\# Online Phase:} 
\State $\mathcal{S}$ computes $e_4 \leftarrow \langle x \rangle_1-b_4$ and sends $e_4$ to $\mathcal{C}$.
\Statex $\mathcal{C}$ computes $f_5 \leftarrow \langle x \rangle_0-a$ and sends $f_5$ to $\mathcal{S}$.
\State $\mathcal{S}$ computes  $ \langle z\rangle_1 \leftarrow p_2(f_5 +\langle x \rangle_1 )^2 +\langle p_2 a^2 \rangle _1 +2((f_5 +\langle x \rangle_1 )(f_4+\langle p_2 a \rangle_1)+\langle c_4\rangle_1) +p_1(f_5 +\langle x \rangle_1 )+\langle p_1 a \rangle _1 + p_0$.
\Statex $\mathcal{C}$ computes $\langle z\rangle_0 \leftarrow \langle p_2 a^2 \rangle _0+ 2(a_4 (e_4 +f_5)+ \langle c_4\rangle_0) +\langle p_1 a \rangle _0  $.
\end{algorithmic}
\vspace*{-0.2\baselineskip}
\end{algorithm}

\textbf{Extension to higher-degree polynomials.} Considering a $d$-degree polynomial $z=\sum_{i=0}^{d}p_ix^i \rightarrow \sum_{i=0}^{d}p_i(f+a)^i$, where $f=x-a$. We can use the binomial theorem \cite{liu2010essence} ($(f+a)^i=\sum_{k=0}^i\binom{i}{k} a^k f^{i-k}$), then we obtain: 
\vspace{-2mm}
\begin{IEEEeqnarray}{rCL}
 z=\sum_{i=0}^{d}p_i(f+a)^i  = \sum_{i=0}^{d} \sum_{k=0}^i \underline{\left(\begin{array}{l}
i \\
k
\end{array}\right)p_i a^k }f^{i-k}. 
\label{eq:binomial}
\vspace{-2mm}
\end{IEEEeqnarray}

The secure computation rationale for Eq.~\ref{eq:binomial} follows the same pattern as that of Eq.~\ref{eq:QuaPol}, with the protocol being divided into offline and online phases.
Likewise, the underlined parts can be computed during the offline phase, where the client learns $\left < {\binom{i}{k}p_i a^k } \right>_0$ and the server learns $\left <{\binom{i}{k} p_i a^k }\right>_1$ $(\forall i\in \{1,2,\cdots,d\}, k\in\{1,2,\cdots,i\})$. In the online phase, the server reconstructs the plaintext $f$ marked by $a$. At the end, only one-round online communication suffices to securely compute the arbitrary degree polynomial $\langle z \rangle =\sum_{i=0}^{d}\langle p_ix^i \rangle$.

\vspace{-0.4\baselineskip}
\subsection{Secure DReLU Protocol}
\label{sec:DReLULayers}
\vspace{-0.4\baselineskip}
Secure DReLU protocol $\mathtt{SDReLU}$ builds upon the FuncSS-based distributed comparison function $\mathcal{F}^{\text{cmp}}_{a,b}$ \cite{boyle2021function,jawalkar2023orca}, that is, $\mathcal{F}^{\text{cmp}}_{a,b}(x)=b$ if $x<a$; otherwise, $\mathcal{F}^{\text{cmp}}_{a,b}(x)=0$.
Recall from Sec.~\ref{sec:Preliminaries}, the two-party $\mathcal{F}^{\text{cmp}}_{a,b}$ involves a pair of algorithms: $\texttt{Gen}^{<}(a,b)$ and $\texttt{Eval}^{<}(\gamma,k_{\gamma},x)$. 
$\texttt{Gen}^{<}(a,b)$ creates two keys $k_0,k_1$, then two parties learn the shared results $\mathcal{F}_{\gamma}^{\text{cmp}} \leftarrow \texttt{Eval}^{<}(\gamma,k_{\gamma},x)$ of $\mathcal{F}^{\text{cmp}}_{a,b}(x)$.
%
There are two challenges in employing $\mathcal{F}^{\text{cmp}}_{a,b}$ for $\mathtt{SDReLU}$. \textit{Challenge 1}: As $x$ is in secret-shared form and needs to remain confidential, calling $\texttt{Eval}^{>}(\gamma,k_{\gamma},x)$ by two parties poses a dilemma. \textit{Challenge 2}: $\mathcal{F}^{\text{cmp}}_{a,b}$ is designed for less-than comparisons, whereas $\sysname$ requires greater-than comparisons, as seen in layers like ReLU's comparison of $x$ against $0$ or maxpool's selection of the maximum value. Therefore, we made modifications to seamlessly integrate FuncSS-based $\mathcal{F}^{\text{cmp}}_{a,b}$ for efficient comparison of secret-shared data in $\sysname$. We present our approach below.

\textit{Addressing challenge 1.} For a signed $l$-bit $x \in \mathbb{Z}_{2^{l}}$, $\textnormal{DReLU}(x)$ is defined as:
\vspace{-1mm}
\begin{IEEEeqnarray}{rCL}
 \textnormal{DReLU}(x)=1\left\{x<2^{l-1}\right\}=1 \oplus \textnormal{MSB}(x). 
 \label{drelu}
 \vspace{-2mm}
\end{IEEEeqnarray}
We first define an offset function $f^a(x) = f(x - a)$ and establish the FuncSS-centric scheme accordingly, where $a \stackrel{\$}{\leftarrow} \mathbb{Z}_{2^{l}}$ is randomly generated by $\mathcal{C}$ in $\mathbb{Z}_{2^l}$. 
Specifically, $\mathcal{C}$ possesses the offset value of the input $x$ and reveals the offset/masked input $x + a \rightarrow \hat{x}$. The FuncSS keys are subsequently computed for $f^a(x + a)$, which is essentially equivalent to evaluating $f(x)$ on $x$.
Accordingly, the offset function of Eq.~\ref{drelu} can be formulated as: 
\vspace{-1mm}
\begin{IEEEeqnarray}{rCL}
\textnormal{DReLU}^{a,c}(\hat{x})&=&\textnormal{DReLU}(\hat{x}-a\pmod {2^l} )=\textnormal{MSB}\left(\hat{x}-{a} \pmod {2^l}\right) \oplus 1  \oplus c. \nonumber
\vspace{-1mm}
\end{IEEEeqnarray}
where $a$ and $c$ are the input and output masks.

\textit{Addressing challenge 2.} Inspired by CrypTFlow2 \cite{rathee2020cryptflow2}, 
for shared $\langle x\rangle_0$ and $\langle x\rangle_1$ with $\langle y\rangle_0 = \langle x\rangle_0 \pmod{2^{l-1}}$ and $\langle y\rangle_1 = \langle x\rangle_1 \pmod{2^{l-1}}$, we reformulate $\textnormal{MSB}(x)$ as: $\textnormal{MSB}(x) =\textnormal{MSB}\left(\langle x\rangle_0\right) \oplus \textnormal{MSB}\left(\langle x\rangle_1\right)  \oplus 1\left\{  2^{l-1} - \langle y\rangle_0-1<\langle y\rangle_1 \right\}.$
Then, we set $\langle x\rangle_0 = \hat{x} = x+a$ and $\langle x\rangle_1 = 2^l-a$, we learn:

\vspace{-3mm}
\begin{small}
\begin{IEEEeqnarray}{rCL}
\textnormal{DReLU}^{a,c}(\hat{x}) = \textnormal{MSB}\left(\hat{x}\right) \oplus 1\left\{  2^{l-1} - \langle y\rangle_0-1<\langle y\rangle_1 \right\}  \oplus   \underline{\textnormal{MSB}\left(2^l-a\right) \oplus 1 \oplus c}.
\label{eq:DReLU}
\end{IEEEeqnarray}
\vspace{-3mm}
\end{small}

\noindent Hence, $\mathtt{SDReLU}$ is divided into online and offline phases, where the computations in the offline phase of Eq.~\ref{eq:DReLU} are labeled with underline as these are independent of the input $x$. Simultaneously, keys are prepared for online $\mathtt{Eval}^{<}$ (i.e., $\mathcal{F}^{\text{cmp}}_{\langle y\rangle_1,1}(2^{l-1} - \langle y\rangle_0-1)$) by invoking $\mathtt{Gen}^{<}$. In the online phase, through only one communication round, $\mathcal{C}$ and $\mathcal{S}$ jointly recover the plaintext $\hat{x}$ for $\mathcal{S}$ learning $\textnormal{MSB}\left(\hat{x}\right)$. Meanwhile, with $\langle y\rangle_0 \leftarrow \hat{x}\pmod {2^{l-1}}$, $\mathcal{C}$ and $\mathcal{S}$ learn the Boolean-shared $1\left\{  2^{l-1} - \langle y\rangle_0-1<\langle y\rangle_1 \right\}$ locally based on the respective key. Algorithm~\ref{alg:drelu} presents the $\mathtt{SDReLU}$ protocol's main steps for executing $\textnormal{DReLU}^{a,c}(\hat{x})$. 

\begin{algorithm}[!t]
\caption{\label{alg:drelu} {Secure DReLU Protocol} }
\textbf{Input:} $\mathcal{C}$ holds $\langle x \rangle_0 \in \mathbb{Z}_{2^{l}}$. $\mathcal{S}$ holds $\langle x \rangle_1 \in \mathbb{Z}_{2^{l}}$.
\par
\textbf{Output:} $\mathcal{C}$ learns $\langle z \rangle_0 \in \mathbb{Z}_{2}$ . $\mathcal{S}$ learns $\langle z \rangle_1 \in \mathbb{Z}_{2}$, where $z = \text{DReLU}(x)$.
\begin{algorithmic}[1]
\Statex \textbf{\# Offline Phase:} $\texttt{Gen}^{\text{DReLU}}$
\State Sample $\langle a \rangle_0,\langle a \rangle_1 \stackrel{\$}{\leftarrow} \mathbb{Z}_{2^{l}},b {\leftarrow} 1$, and
$a=\langle a \rangle_0 + \langle a \rangle_1 \pmod 2^{l}$
\State Let $\langle x\rangle_1 = 2^l- a \in \mathbb{Z}_{2^l}$, $\langle y\rangle_1 = 2^l- a \pmod {2^{l-1}} \in \mathbb{Z}_{2^{l-1}}$.
\State $\{\tilde{k}_0, \tilde{k}_1\} {\leftarrow} \texttt{Gen}^{<}(1^\kappa, \langle y\rangle_1, b)$.
\State Sample a Boolean randomness $c \stackrel{\$}{\leftarrow} \mathbb{Z}_{2}$.
\State Let $r= c \oplus \left \lfloor\frac{\langle x\rangle_1}{2^{l-1}}\right\rfloor \oplus 1$.
\State Sample $[r]_0, [r]_1 {\leftarrow} \mathbb{Z}_{2}$, s.t., $[r]_0 \oplus [r]_1 = r $.
\State $ \forall \gamma \in\{0,1\}, k_{\gamma}= \tilde{k}_{\gamma}||[r]_{\gamma}$.
\State \textbf{Return} $\{\langle a \rangle_{\gamma}, k_{\gamma}\}$

\Statex \textbf{\# Online Phase:} $\texttt{Eval}^{\text{DReLU}}$
\State  $ \forall \gamma \in\{0,1\}$, parse $k_{\gamma}= \tilde{k}_{\gamma}||[r]_{\gamma}$.
\State $\mathcal{C}$ computes $\langle f \rangle_0 \leftarrow \langle x \rangle_0 + \langle a \rangle_0$ and sends $\langle f \rangle_0$ to $\mathcal{S}$.  
\State $\mathcal{S}$ computes $\langle f \rangle_1 \leftarrow \langle x \rangle_1 + \langle a \rangle_1$ and sends $\langle f \rangle_1$ to $\mathcal{C}$. 
\State $\mathcal{C}$ and $\mathcal{S}$ recover $f \leftarrow \langle f \rangle_0
+ \langle f \rangle_1 \pmod {2^{l-1}}$.
\State $ \forall \gamma \in\{0,1\}$, $[\tilde{z}]_{\gamma} \leftarrow \texttt{Eval}^{<}(\gamma, \tilde{k}_{\gamma}, 2^{l-1}-f-1)$.
\State $ \forall \gamma \in\{0,1\}$, $[{z}]_{\gamma} \leftarrow \gamma  \left \lfloor\frac{x+a}{2^{l-1}}\right\rfloor \oplus [r]_{\gamma} \oplus  [\tilde{z}]_{\gamma}$.
\end{algorithmic}
\end{algorithm}

\textbf{Secure ReLU layer.} ReLU($x$) can be expressed as ReLU($x$)$=x \cdot \textnormal{DReLU}(x)$.
To learn the result of secure ReLU layer \textsf{SecReLU} over the secret-shared $\langle x \rangle$, $\mathcal{C}$ and $\mathcal{S}$ learn $[{z}] \leftarrow \mathtt{SDReLU}(\langle x \rangle)$ first and then multiply $[{z}]$ to $\langle x \rangle$: \textsf{SecReLU}$(\langle x \rangle) \leftarrow [{z}]\cdot \langle x \rangle$.
However, direct multiplication of arithmetic shares $\langle x\rangle$ and Boolean shares $[{z}]$ is not feasible due to their calculation with different moduli. 
Various approaches tackle this issue by converting Boolean to arithmetic shares, typically requiring two rounds of online communication (i.e., one for conversion and another for multiplication). Upon investigation, $\mathtt{SBitXA}$ protocol (Algorithm 1 in FssNN \cite{yang2023fssnn}) reduces this to one round, with only $n + 1$ bits of online communication per party. 
In this work, we leverage $\mathtt{SBitXA}$ as a black box to learn $\textsf{SecReLU}(\langle x \rangle) \leftarrow \mathtt{SBitXA}(\langle x \rangle,[{z}])$.

\textbf{Secure max pool layer.} The rationale of secure max pool layer \textsf{SecMaxPool} in NNs is to select the maximum value from $n^2$ shared elements with an $(n\times n)$-width pooling window. Given two elements $\left< x\right>$ and $\left< y\right>$, we reduce \textsf{SecMaxPool} to \textsf{SecReLU} according to below equation:
\vspace{-2mm}
\begin{IEEEeqnarray}{rCL}
 \max (\left< x\right>,\left< y\right>) &=&x \cdot 1\left\{\langle x\rangle > \langle y\rangle \right\}+ y \cdot (1-1\left\{\langle x\rangle > \langle y\rangle \right\}) \nonumber \\
 &=&\underline{(\langle x\rangle -\langle y\rangle) \cdot 1\left\{\langle x\rangle -\langle y\rangle> 0 \right\}} +\langle y\rangle.
 \label{eq:maxpool}
 \vspace{-2mm}
\end{IEEEeqnarray}
The underlined part in Eq.~\ref{eq:maxpool} can be learned by invoking \textsf{SecReLU} protocol. Hence, \textsf{SecMaxPool} can be achieved by invoking \textsf{SecReLU} $(n^2-1)$ times for each $(n\times n)$-width pooling window.

\vspace{-0.6\baselineskip}
\subsection{Secure Piecewise Polynomials Protocol}
\label{sec:SmoothLayers} 
\vspace{-0.6\baselineskip}
Sigmoidal activations in NNs pose challenges in 2PC due to the computational complexity of exponentiation and reciprocal, especially when working with blinded exponents. It is well-known that such activations can be approximated using piecewise continuous polynomials with negligible accuracy loss \cite{liu2017oblivious}. However, existing methods efficiently handle only low-degree polynomials, limiting

\noindent  approximation accuracy.
To address this, we introduce a secure piecewise polynomial protocol $\mathtt{SPiePol}$, that can be computed by our computational- and communication-saving designs.
Given a piecewise polynomial function:

\vspace{-4mm}
\begin{IEEEeqnarray}{rCL}
P(x) = P_1(x) \textbf{1}_{x \in (-\infty, e_1)} +P_2(x) \textbf{1}_{x \in [e_1, e_2)}+ \cdots + P_k(x) \textbf{1}_{x \in [e_{k-1}, \infty)},  
\label{eq:PolyForSigmoidTanh}
\end{IEEEeqnarray}
 
\noindent where $P_i(x) = {p_i}_0 + {p_i}_1 x + \ldots + {p_i}_d x^d$ ($\forall i \in \{1,\cdots, k\}$) is a $d$-degree polynomial applied to different intervals of $x$. 
The indicator function $\textbf{1}_{x \in [e_{i-1},e_i)}$ ensures that each polynomial $P_i(x)$ is active only within its designated interval $[e_{i-1}, e_i)$.  
The $k$ piecewise polynomials in $P(x)$ are transformed into a construction by DReLU (Eq.~\ref{eq:S-Poly1}), where each polynomial $P_i(x)$ is activated by an indicator $s_i$. 
For example, $P_2(x)$ in Eq.~\ref{eq:PolyForSigmoidTanh} is activated when $e_1 \leq x < e_2$, i.e., $s_2=1$ and all other indicators are ``0''.
The indicator $s_2$ is derived from two DReLUs: $\neg {c}_1 \leftarrow  \neg \text{DReLU}(e_1 -x)=1$, indicating $e_1 \leq x$, and ${c}_2 \leftarrow \text{DReLU}(e_2-x)=1$, indicating $x < e_2$, with $s_2 = \neg{c}_1 \otimes {c_2}$, where $\neg$ and $\otimes$ represent logical NOT and AND, respectively. 
Thus, $P(x)$'s expression is given by:
\vspace{-2mm}
{\begin{IEEEeqnarray}{rCL}
P(x) &=&  \text{DReLU}(e_1 - x) \cdot P_1(x)   + \sum_{i=1}^{k-2} \neg \text{DReLU}(e_i-x) \cdot \text{DReLU}(e_{i+1} - x) \label{eq:S-Poly1} \nonumber \\
&& \cdot P_{i+1}(x) + \neg \text{DReLU}(e_{k-1}-x) \cdot P_k(x). 
\label{eq:S-Poly2}
\end{IEEEeqnarray}}

To this end, we present the secure realization of $\mathtt{SPiePol}$ over the secret-sharing domain based on Eq.~\ref{eq:S-Poly1}. 
Given the additive shares of each neuron output $\langle x \rangle$, and ${p_i}_j (i \in [k], j\in [d])$ representing the plaintext weights held by the server $\mathcal{S}$, the client $\mathcal{C}$ and server $\mathcal{S}$ collaboratively compute $\langle z \rangle \leftarrow P(\langle x \rangle) = \mathtt{SPiePol}(\langle x \rangle)$. The main steps are outlined in Algorithm~\ref{alg:Smooth}.

\begin{algorithm}[!t]
\caption{\label{alg:Smooth} {Secure Piecewise Polynomials Protocol} }
\textbf{Input:} $\mathcal{C}$ holds $\langle x \rangle_0 \in \mathbb{Z}_{2^{l}}$. $\mathcal{S}$ holds $\langle x \rangle_1 \in \mathbb{Z}_{2^{l}}$.
\par
\textbf{Output:} $\mathcal{C}$ learns $\langle z \rangle_0 \in \mathbb{Z}_{2}$ . $\mathcal{S}$ learns $\langle z \rangle_1 \in \mathbb{Z}_{2}$, where $z = P(x)$.
\begin{algorithmic}[1]
\State $\mathcal{C}$ and $\mathcal{S}$ execute $\mathtt{SDReLU}$ $k-1$ times, resulting in $k-1$ shared comparison bits: $\forall i\in \{1,2,\cdots, k-1\}$, $[ c_i ] \leftarrow \mathtt{SDReLU}(e_i - \langle x\rangle)$. 
\State $\mathcal{C}$ and $\mathcal{S}$ execute $\mathtt{SQuaPol}$ $k$ times, resulting in $k$ shared polynomial results: $\forall i\in \{1,2,\cdots, k\}$, $\langle f_i \rangle \leftarrow \mathtt{SQuaPol}(P_i(\langle x \rangle))$. 
\State $\mathcal{C}$ and $\mathcal{S}$ compute the $1^{\textnormal{st}}$-th polynomial via $\left\langle z_1\right\rangle=\mathtt{SBitXA}([ c_1], \langle f_1 \rangle)$. 
\For{$1<i<k$}
\State $\mathcal{C}$ and $\mathcal{S}$ compute the $i$-th polynomial via $\left\langle z_i\right\rangle=\mathtt{SBitXA}([c_{i-1}] \oplus \gamma, \mathtt{SBitXA}([ c_i], \langle f_i \rangle))$. 
\EndFor
\State $\mathcal{C}$ and $\mathcal{S}$ compute the $k$-th polynomial via $\left\langle z_k\right\rangle=\mathtt{SBitXA}([ c_{k-1}] \oplus \gamma, \langle f_k \rangle)$. 
\State $\mathcal{C}$ and $\mathcal{S}$ learn the AddSS-shared $P(\langle x\rangle)$: $\langle z \rangle \leftarrow \sum_{i=1}^{k}\langle z_i \rangle$. 
\end{algorithmic}
\end{algorithm}

\textbf{Secure Sigmoid and Tanh layers.} 
The server offers an efficient alternative to the expensive exponentiation and reciprocal operations inherent in standard sigmoidal activations, i.e., Sigmoid and Tanh. 
By approximating these activations with piecewise continuous polynomials (e.g., splines \cite{dierckx1995curve}), the server learns the plaintext polynomial weights. 
As a result, secure Sigmoid and Tanh layers i.e., $\textsf{SecSig}$ and $\textsf{SecTanh}$, can be implemented directly using $\mathtt{SPiePol}$.

\vspace{-0.4\baselineskip}
\subsection{Putting Things Together}
\label{sec:PrivInf}
We are ready to integrate these blocks into the scheme $\sysname$. $\sysname$ considers the most widely used GNN model for graph classification tasks, namely message passing neural networks (MPNNs).
The client $\mathcal{C}$ first splits the graph-structured data $D = (G, E, H)$, where $G \in \mathbb{Z}_2^{n \times n}$ is the adjacency matrix, $E \in \mathbb{R}^{n \times n}$ represents the attributes of the edges, and $H \in \mathbb{R}^{n \times m}$ is the node feature matrix, with $n$ as the number of nodes and $m$ as the feature dimension. The data is split into random shared values ${\langle D \rangle_0, \langle D \rangle_1}$. At this point, the structure and data of the graph are fully protected.
$\mathcal{C}$ then uploads $\langle D \rangle_1= (\langle G \rangle_1,\langle E\rangle_1,\langle H \rangle_1)$ to the server to access $\sysname$ services. 
%
Due to space limitations, we outline the secure inference (SI) process in MPNN-centric $\sysname$. 
The SI process involves three functions: secure message passing function (\textsc{PrivMF}), secure update function (\textsc{PrivUF}), and secure readout function (\textsc{PrivRF}) --- which are securely executed as detailed below. 
 
\begin{itemize}[leftmargin=*,topsep=0pt]
\item[-] \textsc{PrivMF} consists of two types of NN layers: $\textsf{SecFC}$ and $\textsf{SecReLU}$ for message passing. With GS data $\langle D \rangle$ and server-hold plaintext parameter $W$, \textsc{PrivMF}'s workflow is $(\langle D\rangle,W) \rightarrow [\textsf{SecFC} \rightarrow \textsf{SecReLU}] \times \iota  \rightarrow \textsf{SecFC} \rightarrow  \mathtt{AGGREGATE}(\cdot)  \rightarrow \langle M \rangle=\{\langle m_v \rangle |v\in V \}=  \{\langle \sum_{u \in N(v)} m_{u \rightarrow v} \rangle |v\in V\}$, where $\iota$ denotes the number of repetitions, $\mathtt{AGGREGATE}$ represents the aggregation function, $V$ is the set of nodes in graph, and $N(v)$ is the set of neighboring nodes of node $v$.  
\item[-] \textsc{PrivUF} sets up secure update gates and secure reset gates, using \textsf{SecSig}, \textsf{SecTanh}, and $\mathtt{SEleMul}$ to compute the update states of the two gates. For each node $v\in V$ (i.e., $m_v\in M$ and $h_v\in H$), \textsc{PrivUF}'s workflow is 
$(\langle m_v\rangle,\langle h_v\rangle, W^0)$ $\rightarrow [\textsf{SecFC} \rightarrow \textsf{SecReLU}]+[\textsf{SecFC} \rightarrow \textsf{SecReLU}] \rightarrow \textsf{SecSig} \rightarrow \langle \varphi_v \rangle$, 
$(\langle m_v\rangle,\langle h_v\rangle, W^1)$ $\rightarrow [\textsf{SecFC} \rightarrow \textsf{SecReLU}]+[\textsf{SecFC} \rightarrow \textsf{SecReLU}] \rightarrow \textsf{SecSig} \rightarrow \langle \eta_v\rangle$, 
$(\langle m_v\rangle,\langle h_v\rangle, W^2)$ $\rightarrow [\textsf{SecFC} \rightarrow \textsf{SecReLU}] \odot \langle \eta_v\rangle + [\textsf{SecFC} \rightarrow \textsf{SecReLU}] \rightarrow \textsf{SecTanh} \rightarrow \langle \vartheta_v\rangle$, and $(\langle \varphi_v\rangle, \langle \vartheta_v\rangle,\langle h_v\rangle) \rightarrow (1-\langle \varphi_v\rangle ) \odot \vartheta_v + \langle \varphi_v\rangle  \odot \langle h_v\rangle  \rightarrow \langle\hat{h}_v\rangle$. The overall updated node feature matrix is $\langle \hat{H}\rangle=\{\langle\hat{h}_v\rangle|v\in V\}$. 
\item[-] With updated $\hat{H}^{(1)}$ and $\hat{H}^{(T)}$, where $\hat{H}^{(t)}$ denotes the outputs of $t$ executions of $\textsc{PrivMF}$ $\overset{\rightarrow  }{\leftarrow}$ $\textsc{PrivUF}$, \textsc{PrivRF} executes two paths: $(\langle \hat{H}^{(1)}\rangle |\langle \hat{H}^{(T)}\rangle ,W_R) \rightarrow [\textsf{SecFC} \rightarrow \textsf{SecReLU}] \times \iota  \rightarrow \textsf{SecFC} \rightarrow \langle \tilde{\mathcal{HH}}^{\iota+1} \rangle$ and $(\langle \hat{H}^{(T)}\rangle ,W_Z) \rightarrow [\textsf{SecFC} \rightarrow \textsf{SecReLU}] \times \iota  \rightarrow \textsf{SecFC} \rightarrow \langle  \tilde{\mathcal{H}}^{\iota+1} \rangle$, where $\iota$ denotes the number of repetitions and $|$ denotes concatenation. \textsc{PrivRF} then connect the results of the two paths: $\langle \mathcal{R}\rangle= \mathtt{SMatMul}(\mathtt{SEleMul} ( \textsf{SecSig}(\langle  \tilde{\mathcal{HH}}^{\iota+1}  \rangle), \langle \tilde{\mathcal{H}}^{\iota+1} \rangle)$, $\mathtt{SDReLU}(\langle \tilde{\mathcal{H}}^{\iota+1} \rangle)$.
\end{itemize}

\vspace{-0.4\baselineskip}
\section{Theoretical Analysis}
\label{sec:TheoreticalAnalysis}
\vspace{-0.6\baselineskip}

\begin{table*}[!t] 
\caption{Computation and communication costs of supporting protocols in $\sysname$, PAPI \cite{cheng2023private}, and FastSecNet \cite{hao2023fastsecnet}.}
\vspace{-3mm}
\centering
\begin{adjustbox}{width=1\textwidth,center}
\begin{tabular}{cccc}
 \hline 
{Protocols} & Benchmarks & Computation & Communication \\ \hline 
\multirow{2}{*}{$\mathtt{SMatMul}$} & PAPI & $\mathcal{T}_{\text{SMul}}$ & $2l$ (online: $2l$)\\  
&  $\sysname$    & $\mathcal{T}_{\text{SMul}}$ & $2l$ (online: $l$)\\ \hline 
\multirow{2}{*}{$\mathtt{SQuaPol}$} & PAPI    &$7\mathcal{T}_{\text{HEnc}}+ 3\mathcal{T}_{\text{HDec}}+7\mathcal{T}_{\text{HAdd}}+5\mathcal{T}_{\text{HMul}}+ \mathcal{T}_{\text{SMul}}$ & $5l_h+2l$ (online: $2l$) \\  
&  $\sysname$    &$\mathcal{T}_{\text{SMul}}$ & $9l$ (online: $2l$) \\ \hline  
\multirow{2}{*}{$\mathtt{SDReLU}$}&  FastSecNet    & $2\mathcal{T}_{\text{SFunC}}$ & $8(l+1)$ (online: $2l$)  \\ 
&  $\sysname$  & $\mathcal{T}_{\text{SFunC}}+\mathcal{T}_{\text{SMul}}$  & $2(2l+1)$ (online: $2l$)  \\ \hline  
\multirow{2}{*}{$\mathtt{SPiePol}$}&  FastSecNet    & $2(k-1)\mathcal{T}_{\text{SFunC}}+ \frac{d(d+1)k}{2}\mathcal{T}_{\text{SMul}}$& $8(k-1)(l+1)+d(d+1)kl$ (online: $2l(k-1)+\frac{d(d+1)kl}{2}$)  \\ 
&  $\sysname$ &$(k-1)\mathcal{T}_{\text{SFunC}} + (2k-1)\mathcal{T}_{\text{SMul}}$ & $6kl+2k-4l-2$ (online: $4kl-2l$) \\ \hline 
\end{tabular}
\end{adjustbox} 
\vspace{-2mm}
\label{tab:comm}
\end{table*}

\textbf{Efficiency analysis.}
We analyze the computational and communication complexity of key protocols ($\mathtt{SMatMul}$, $\mathtt{SQuaPol}$, $\mathtt{SDReLU}$, and $\mathtt{SPiePol}$) and compare them with SotA schemes.
We define {SMul} and {SFunC} as AddSS-based multiplication and FuncSS-based comparison operations. {HEnc}, {HDec}, {HAdd}, and {Hmul} represent the operations for encryption, decryption, addition, and multiplication with homomorphic ciphertexts.
There are five parameters involved in this analysis: (1) the bit length $l$ of an AddSS share, (2) the bit $l_h$ length of the homomorphic ciphertext, (3) the pieces $k$ of piecewise polynomials, (4) the degree $d$ of each piece, (5) the time complexity $\mathcal{T}$ of secure operations.  
Table~\ref{tab:comm} summarizes the analysis results.
$\mathtt{SQuaPol}$, $\mathtt{SDReLU}$, and $\mathtt{SPiePol}$ show notable overall computational improvements, with most costs concentrated in the offline phase for generating cryptographic primitives like multiplication triples and keys. They also achieve lower overall communication costs compared to SotA. 

\vspace{2mm}
\textbf{Security analysis.} 
\sysname's pipeline integrates a variety of cryptographic protocols for different layers, with each layer's input and output in the additive secret-sharing domain. Using the sequential composition theorem \cite{goldreich2019play}, we deduce the overall security of \sysname's inference as stated in Theorem~\ref{theo}.

\begin{theorem}
\label{theo}
$\sysname$'s secure inference scheme $\prod^{\sysname}$ securely realizes the ideal functionality $\mathcal{F}^{\sysname}$ in the presence of one semi-honest adversary $\mathcal{A}$ in the \textnormal{(}$\prod_{\textsf{SecCONV}}$, $\prod_{\textsf{SecReLU}}$, $\prod_{\textsf{SecMaxPool}}$, $\prod_{\textsf{SecSig}}$, $\prod_{\textsf{SecTanh}}$,$\prod_{\textsf{SecFC}}$, $\prod_{\mathtt{SEleMul}}$ \textnormal{)}-hybrid model.
\end{theorem}

We analyze the security of $\sysname$ against two types of semi-honest adversaries, as described in \textbf{Definition}~\ref{security}. Specifically, we consider the following two cases based on the potential adversary: (1) a corrupted client ($\mathcal{P} \leftarrow \mathcal{C}$) and (2) a corrupted server ($\mathcal{P} \leftarrow \mathcal{S}$).
The security of $\sysname$ under these two semi-honest adversaries is proven according to the cryptographic standard outlined in \textbf{Definition}~\ref{security}.
%


\vspace{-0.4\baselineskip}
\section{Experimental Evaluation}
\label{sec:ExperimentalEvaluation}  
\vspace{-0.6\baselineskip}
\textbf{Testbed.} We implement a prototype of $\sysname$ in Python 3.7 and PyTorch 1.9. Extensive experiments are conducted on two distinct servers with 64-core CPUs, 128GB RAM, and 2 NVIDIA GeForce RTX 2080Ti GPU. The environment simulates a local-area network with 1 Gbps bandwidth and 0.1 ms latency. Following prior work \cite{riazi2019xonn,liu2021towards,hao2023fastsecnet}, we set the integer ring size of additive secret shares to $\mathbb{Z}_{2^{32}}$.
Cleartext NN models are trained using PyTorch on an NVIDIA RTX 2080Ti GPU, using the standard SGD optimizer with a learning rate of $0.001$, batch size of 128, momentum of 0.9, and weight decay of $1 \times 10^{-6}$.

{\textbf{Datasets and models.} We evaluate $\sysname$ on MNIST, CIFAR-10, CIFAR-100, and QM9 \cite{ramakrishnan2014quantum} datasets. 
%
MNIST, CIFAR-10, and CIFAR-100, are utilized for image classification inference with CNNs. 
The QM9 dataset, comprising 134,000 stable organic molecules, is processed into molecular fingerprints using RDKit \cite{rdkit}, and employed for molecular property inference with GNNs.
The models include a 3-layer fully-connected network (FC-3) and a 4-layer CNN (CNN-4) \cite{riazi2019xonn} for MNIST, LeNet \cite{lecun1998gradient} and VGG-16 \cite{simonyan2014very} for CIFAR10, ResNet-32 \cite{he2016deep} and VGG-16 \cite{simonyan2014very} for CIFAR-100, and the GNN in Sec.~\ref{sec:PrivInf} for QM9.}

\vspace{-0.4\baselineskip}
\subsection{Microbenchmarks}
\vspace{-0.4\baselineskip}
\label{ssec:SupportingProtocols} 
\begin{figure*}[!t]
    \centering
    \begin{minipage}[t]{0.24\linewidth}
        \centering
    \includegraphics[width=\textwidth]{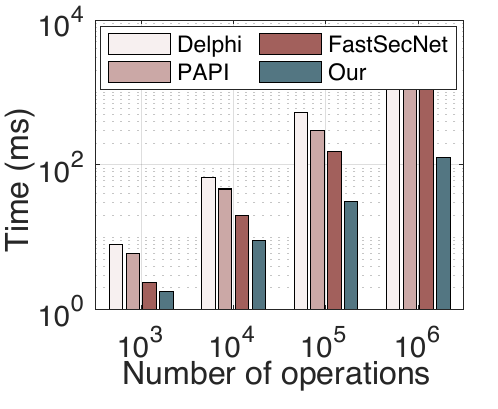}
      \centerline{\footnotesize{(a) Offline runtime}}\medskip
        \vspace{-1mm} 
    \end{minipage}%
    \hfill
    \begin{minipage}[t]{0.24\linewidth}
        \centering
    \includegraphics[width=\textwidth]{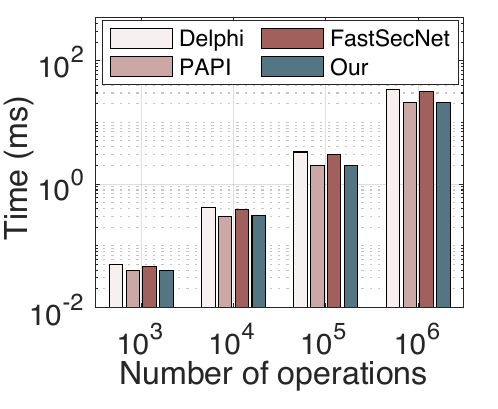}
        \centerline{\footnotesize{(b)} Online runtime }\medskip
        \vspace{-1mm}
    \end{minipage}
     \hfill
    \begin{minipage}[t]{0.24\linewidth}
        \centering
    \includegraphics[width=\textwidth]{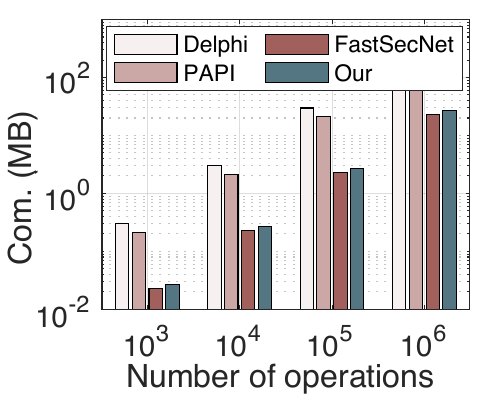}
          \centerline{\footnotesize{(c)} Offline com. cost}\medskip
        \vspace{-1mm}
    \end{minipage}
     \hfill
    \begin{minipage}[t]{0.24\linewidth}
        \centering
    \includegraphics[width=\textwidth]{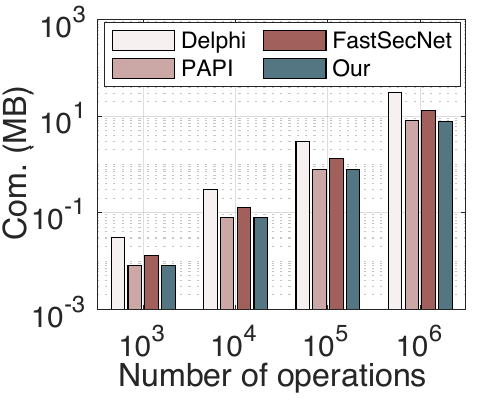}
          \centerline{\footnotesize{(d)} Online com. cost }\medskip
        \vspace{-1mm}
    \end{minipage}
    \vspace{-0.8\baselineskip}
     \caption{Performance comparison of $\mathtt{SQuaPol}$ protocol.} 
        \label{fig:quapol}
        \vspace{-2mm}
\end{figure*}

\textbf{Secure $\mathtt{SQuaPol}$ protocol}. Fig.~\ref{fig:quapol} compares the performance of $\mathtt{SQuaPol}$ with Delphi \cite{mishra2020delphi}, PAPI \cite{cheng2023private}, and FastSecNet \cite{hao2023fastsecnet} in terms of runtime and communication (com.) cost across operation scales from $10^3$ to $10^6$.
$\mathtt{SQuaPol}$ outperforms in online computation, showing a $\sim 35\%$ reduction in time compared to Delphi and FastSecNet, benefiting from a single-round interaction. Online com. of $\mathtt{SQuaPol}$ also excels, with costs $4 \times$ and $2 \times$ lower than Delphi and FastSecNet. 
For offline computation, $\mathtt{SQuaPol}$ achieves improvements by one order of magnitude over other schemes. 
While its offline com. cost is $\sim 20\%$ higher than FastSecNet, $\mathtt{SQuaPol}$ prioritizes online efficiency, delivering strong overall performance. 

\begin{figure*}[!t]
    \centering
    \begin{minipage}[t]{0.24\linewidth}
        \centering
    \includegraphics[width=\textwidth]{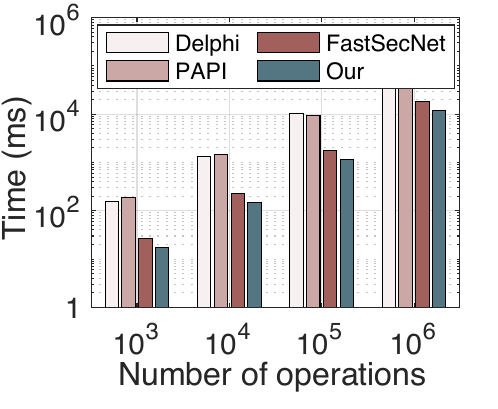}
      \centerline{\footnotesize{(a) Offline runtime}}\medskip
        \vspace{-1mm} 
    \end{minipage}%
    \hfill
    \begin{minipage}[t]{0.24\linewidth}
        \centering
    \includegraphics[width=\textwidth]{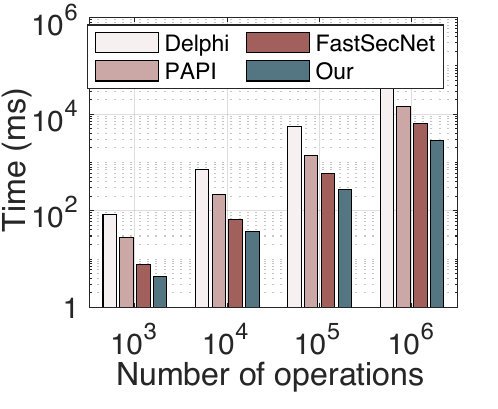}
          \centerline{\footnotesize{(b)} Online runtime }\medskip
        \vspace{-1mm}
    \end{minipage}
     \hfill
    \begin{minipage}[t]{0.24\linewidth}
        \centering
    \includegraphics[width=\textwidth]{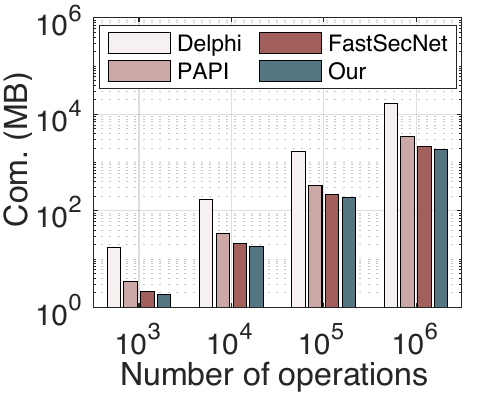}
          \centerline{\footnotesize{(c)} Offline com. cost}\medskip
        \vspace{-1mm}
    \end{minipage}
     \hfill
    \begin{minipage}[t]{0.24\linewidth}
        \centering
    \includegraphics[width=\textwidth]{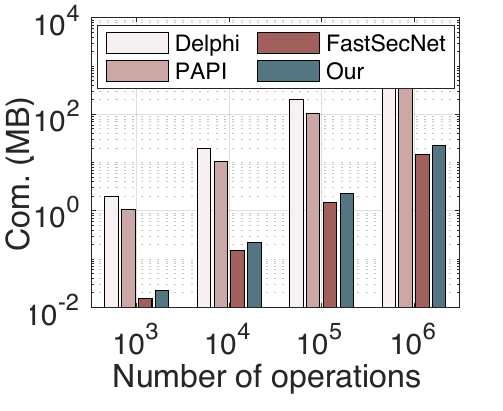}
          \centerline{\footnotesize{(d)} Online com. cost }\medskip
        \vspace{-1mm}
    \end{minipage}
    \vspace{-0.8\baselineskip}
     \caption{Performance comparison of \textsf{SecReLU} protocol.} 
        \label{fig:ReLU}
        \vspace{-0.2\baselineskip}
\end{figure*}

\textbf{Secure comparison-based activations.}
Fig.~\ref{fig:ReLU} reports the performance comparison of \textsf{SecReLU}. 
\textsf{SecReLU} achieves $2\times \sim 15\times$ speedup in online runtime compared to Delphi, PAPI, and FastSecNet, with higher online com. costs than FastSecNet. 
Specifically, \textsf{SecReLU} requires one FuncSS-based comparison and one multiplication for $\text{ReLU}(x) = \text{DReLU}(x) \cdot x$, while FastSecNet uses two comparisons, trading online time for lower com. In contrast, Delphi and PAPI's use of garbled circuits and HE introduces higher overall costs. 
Further, \textsf{SecReLU} presents superior offline performance in both runtime and com. costs, stressing its overall efficiency.


\begin{figure}[!t]
    \centering
    \begin{minipage}[c]{0.47\linewidth}  
    \vspace{-5mm}
        \caption{Performance comparison of \textsf{SecSig}/\textsf{SecTanh} protocols: SiRNN uses a 3-piece polynomial, MiniONN a 12-piece linear approximation, while our protocol employs a 12-piece quadratic approximation, offering better accuracy due to the higher polynomial degree.}
        \label{fig:Sigmoid}
    \end{minipage}%
    \hfill
    \begin{minipage}[c]{0.5\linewidth}  
        \centering
        \begin{minipage}[t]{0.48\linewidth}  
            \centering
            \includegraphics[width=\textwidth]{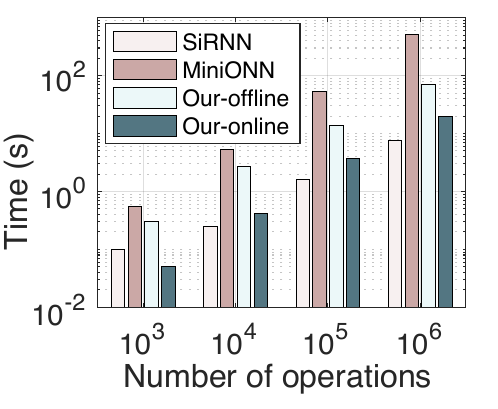}
            \centerline{\footnotesize{(a) Runtime}}\medskip
        \end{minipage}%
        \hfill
        \begin{minipage}[t]{0.48\linewidth}  
            \centering
            \includegraphics[width=\textwidth]{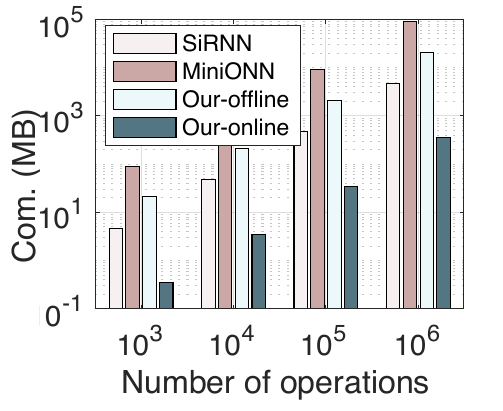}
            \centerline{\footnotesize{(b) Com. cost}}\medskip
        \end{minipage}
    \end{minipage}
    \vspace{-1\baselineskip}
\end{figure}

\textbf{Secure sigmoidal activations.} 
Fig.~\ref{fig:Sigmoid} reports the performance of \textsf{SecSig} (equivalent to \textsf{SecTanh} and $\mathtt{SPiePol}$) compared to SiRNN \cite{rathee2021sirnn} and MiniONN \cite{liu2017oblivious}. 
\textsf{SecSig} adopts a 12-piece spline, sufficient for maintaining cross-entropy loss \cite{liu2017oblivious}.  
We compare its online portion to the total costs of the fully online baselines.  
The online running time and com. cost of \textsf{SecSig} are both less than those of SiRNN and MiniONN for small scales. For $10^3$ operations, the online running time of \textsf{SecSig} is $2 \times$ and $10.8 \times$ faster than SiRNN and MiniONN. The 3-piece spline of SiRNN shows sub-linear growth in latency but offers limited approximation accuracy. 
Regarding online com. cost, \textsf{SecSig} is $13.6 \times$ and $260 \times$ lower than SiRNN and MiniONN across all operation scales. 

\vspace{-0.4\baselineskip}
\subsection{Summary of Accuracy }
\label{ssec:SummaryOfAccuracy}\vspace{-0.4\baselineskip}

\begin{table}[!t] 
\caption{Summary of inference accuracy (\%)}
\vspace{-3mm}
\centering
\begin{adjustbox}{width=0.75\textwidth,center}
\begin{tabular}{lcc|cc|cc|c}
\hline  \hline  
 & \multicolumn{2}{c|}{ MNIST}   & \multicolumn{2}{c|}{ CIFAR-10}   &\multicolumn{2}{c|}{ CIFAR-100}  & QM9 \\ 
 & FC-3 & CNN-4 & LeNet-5 & VGG-16 & ResNet-32 & VGG-16 & GNN \\ \hline
Training Accuracy & 96.96 & 99.16 & 96.81 & 88.03 & 74.90 & 72.22 & 98.17 \\ 
Plain Inference & 96.23 & 99.02 & 81.50 & 87.45 & 68.48 & 70.12 & 97.93 \\ 
Secure Inference & 96.12 & 99.00 & 80.57 & 87.40 & 68.16 & 70.12 & 96.42 \\ \hline\hline  
\end{tabular}
\end{adjustbox}
\label{tab:Accuracy}
\vspace{-2mm}
\end{table}

Table~\ref{tab:Accuracy} compares the inference accuracy of $\sysname$ with plaintext inference across various datasets and models, delivering its robust SI capabilities.
Specifically, for MNIST, $\sysname$ achieves an inference accuracies of $96.12\%$ and $99.00\%$ for the FC-3 and CNN-4 models, respectively, only $\sim 0.1\%$ lower than their plaintext. 
On CIFAR-10, $\sysname$ records accuracies of 80.57\% for LeNet-5 and 87.40\% for VGG-16, slightly below the plaintext accuracies of 81.50\% and 87.45\%, respectively, but still reasonable.
For CIFAR-100, ResNet-32 achieves a SI accuracy of 68.16\%, closely trailing the plaintext accuracy of 68.48\%. VGG-16 matches the plaintext accuracy of 70.12\%.
On QM9 with the GNN model, $\sysname$ reports an accuracy of 96.42\%, slightly below the plaintext accuracy of 97.93\%. The possible minor accuracy drop is from truncation errors caused by the fixed-point arithmetic and the reformulated activations used in $\sysname$.
%

\vspace{-0.4\baselineskip}
\subsection{Performance Comparison with SotA Schemes}
\label{ssec:PerformanceComparison}
\vspace{-0.4\baselineskip}
{\textbf{Secure convolutional inference.} We evaluate $\sysname$ on CNN-4 (MNIST), VGG-16 (CIFAR-10), and ResNet-32 (CIFAR-100), and compare it with SotA works. 
The reported results of SotA are sourced directly from the respective papers.
Table~\ref{tab:InferenceCMP} shows that $\sysname$ improves CNN-4's online runtime up to $1.2\times \sim 73.6\times$.
For VGG-16, $\sysname$ exhibits 2.6$\times$ and 4.0 $\times$ savings in online time when compared to the CrypTFlow2 \cite{rathee2020cryptflow2} and Delphi \cite{mishra2020delphi}, respectively. While $\sysname$'s online runtime is $21.5\%$ higher than PAPI \cite{cheng2023private} due to PAPI's optimization strategy of retaining a subset of ReLUs.
For ResNet-32, $\sysname$ outperforms prior works with online runtime reductions of $1.3 \times \sim 4.7 \times$, also enhancing online communication efficiency by $4.6 \times \sim 23.9 \times$. 

\begin{table}[!t] 
\caption{Performance comparison with SotA schemes (Scheme abbreviations: Medi = MediSC\cite{liu2021towards}, Fast = FastSecNet\cite{hao2023fastsecnet}, Cryp = CrypTFlow2\cite{rathee2020cryptflow2}, Delp = Delphi\cite{mishra2020delphi}, Chet = Cheetah\cite{huang2022cheetah}, SecM = SecMPNN\cite{liao2022secmpnn}).
}
\vspace{-3mm}
\centering
\begin{adjustbox}{width=1\textwidth,center}
\begin{tabular}{clcccc|cccc|cccc|cc}
\hline \hline
\multicolumn{2}{c}{Model}& \multicolumn{4}{c|}{MNIST (CNN-4)}  & \multicolumn{4}{c|}{CIFAR-10 (VGG-16)}   & \multicolumn{4}{c|}{CIFAR-100 (ResNet-32)}  & \multicolumn{2}{c}{QM9 (MPNN)}      \\
\multicolumn{2}{c}{Scheme}            & \multicolumn{1}{l}{XONN} & \multicolumn{1}{l}{Medi} & \multicolumn{1}{l}{Fast} & \multicolumn{1}{l|}{Ours} & \multicolumn{1}{l}{Cryp} & \multicolumn{1}{l}{Delp} & \multicolumn{1}{l}{PAPI} & \multicolumn{1}{l|}{Ours} & \multicolumn{1}{l}{Cryp} & \multicolumn{1}{l}{Chet} & \multicolumn{1}{l}{PAPI} & \multicolumn{1}{l|}{Ours} & \multicolumn{1}{l}{SecM} & \multicolumn{1}{l}{Ours} \\ \hline 
\multirow{3}{*}{\begin{tabular}[c]{@{}c@{}}{Time} \\  {(s)}\end{tabular}}   & Offline & -           & 1.21          & 0.25 & 0.22        & 42.50 & 88.10          & 47.80        & 23.15        & 62.50 & - & 65.70        & 38.36       & - & 5.45        \\ 
 & Online  & 0.30         & 4.42          & 0.07 & 0.06        & 4.20  & 6.30           & 1.30         & 1.53        & 6.40  & 15.95          & 4.50         & 3.35        &    102.57       & 2.48        \\  
 & Total   & 0.30         & 5.63          & 0.32 & 0.28        & 46.70 & 94.40          & 49.10        & 24.68      & 68.90 & 15.95          & 70.20        & 41.71       & 102.57      & 7.93        \\ 
\multirow{3}{*}{\begin{tabular}[c]{@{}c@{}}{Com.} \\  {(MB)}\end{tabular}} & Offline & -           & 2.54          & 40.19& 22.59       & 30.72& 51.20          & 40.96       & 217.74      & 122.88            & - & 143.36      & 468.26      & - & 87.04       \\
 & Online  & 62.77       & 2.62          & 0.65 & 1.28        & 686.08            & 40.96         & 30.72       & 16.55       & 583.68            & 112.64         & 122.88      & 24.38       & 156.13         & 9.10         \\
  & Total   & 62.77       & 5.16          & 40.84& 23.87       & 716.8& 92.16         & 71.68       & 234.29      & 706.56            & 112.64         & 266.24      & 492.64      & 156.13         & 96.14\\\hline \hline
\end{tabular}
\end{adjustbox}
\label{tab:InferenceCMP}
\end{table}

\begin{figure*}[!t]
    \centering
    \begin{minipage}[t]{0.161\linewidth}
        \centering
    \includegraphics[width=\textwidth]{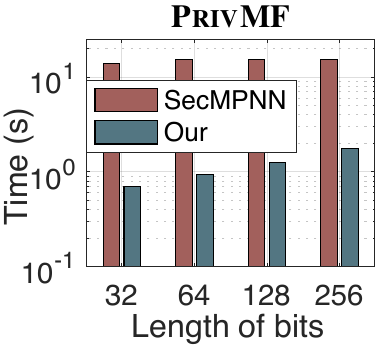}
      \centerline{\footnotesize{(a) Runtime}}\medskip
        \vspace{-1mm} 
    \end{minipage}%
    \hfill
    \begin{minipage}[t]{0.165\linewidth}
        \centering
    \includegraphics[width=\textwidth]{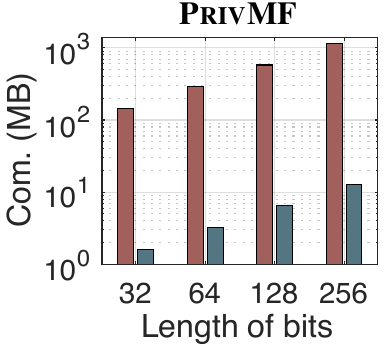}
          \centerline{\footnotesize{(b)} Com. cost }\medskip
        \vspace{-1mm}
    \end{minipage}
    \hfill
    \begin{minipage}[t]{0.163\linewidth}
        \centering
    \includegraphics[width=\textwidth]{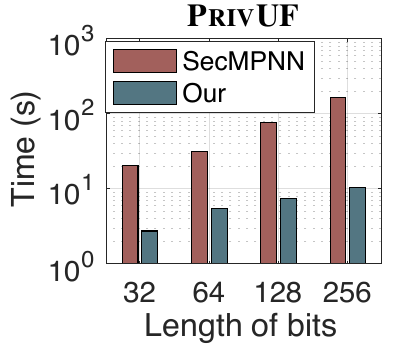}
      \centerline{\footnotesize{(c) Runtime}}\medskip
        \vspace{-1mm} 
    \end{minipage}%
    \hfill
    \begin{minipage}[t]{0.163\linewidth}
        \centering
    \includegraphics[width=\textwidth]{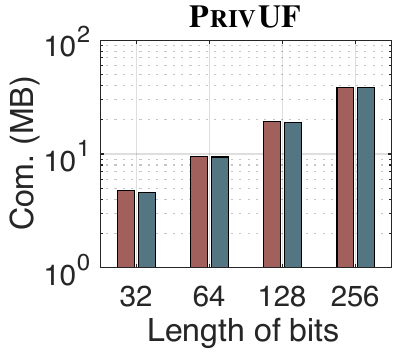}
    \centerline{\footnotesize{(d)} Com. cost }\medskip
        \vspace{-1mm}
    \end{minipage}
    \begin{minipage}[t]{0.163\linewidth}
        \centering
    \includegraphics[width=\textwidth]{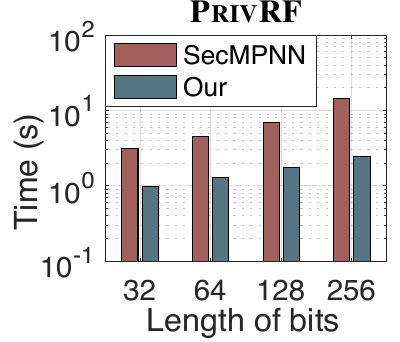}
      \centerline{\footnotesize{(e) Runtime}}\medskip
        \vspace{-1mm} 
    \end{minipage}%
    \hfill
    \begin{minipage}[t]{0.161\linewidth}
        \centering
    \includegraphics[width=\textwidth]{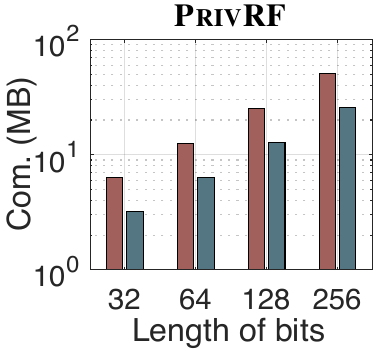}
          \centerline{\footnotesize{(f)} Com. cost }\medskip
        \vspace{-1mm}
    \end{minipage}
    \vspace{-1\baselineskip}
     \caption{Performance comparison of different functions in GNNs. } 
       \vspace{-1.2\baselineskip}
        \label{fig:MPNNComp}
\end{figure*}

{\textbf{Secure graph inference.} 
The overall performance is shown in the last column of Table~\ref{tab:InferenceCMP}. We evaluate three functions across four secret-sharing ring sizes ($l=$ 32, 64, 128, 256 bits) to assess resource usage.  
Fig.~\ref{fig:MPNNComp} shows that time and com. costs rise with larger secret-sharing ring sizes. 
As seen, the three functions \textsc{PrivMF}, \textsc{PrivUF}, and \textsc{PrivRF} take {$0.7$, $2.7$, and $1.0$} seconds, respectively, to predict one molecule within the 32-bit ring. Overall, $\sysname$ achieves an order of magnitude improvement over SecMPNN \cite{liao2022secmpnn}, attributed to the dual optimization of linear and non-linear layers.
Notably, \textsc{PrivUF} accounts for $50\% \uparrow$ of the overall SI runtime due to the costly piecewise approximations of sigmoidal activations like \textsf{SecSig} and \textsf{SecTanh}.

\vspace{-0.4\baselineskip}
\section{Conclusion}
\label{sec:Conclusion}
\vspace{-0.6\baselineskip}
We introduced $\sysname$, a high-performance cryptographic inference framework designed for privacy-preserving GNNs in the semi-honest client-server setup. 
By carefully designing a line of secure offline-online protocols for both linear and non-linear layers with lightweight AddSS and FuncSS, $\sysname$ performs fast and low-interactive inference during the online phase.
Extensive experimental results on benchmark and real-world datasets shed light on the scalability and efficiency of $\sysname$, outperforming prior PPDL and secure GNN schemes. Our approach paves the way for secure, scalable AI-driven services in various domains, including drug discovery and beyond.
%



\newpage
\bibliographystyle{splncs04}
\bibliography{refs}

\end{document}